\documentclass[usenatbib,useAMS,fleqn]{mn2e} 
\pdfoutput=1
\usepackage[usenames,dvipsnames]{color}
\usepackage{graphicx}
\usepackage{amsmath}

\newcommand{\ri}{\rmn{i}}

\renewcommand{\k}{{\bmath{k}}}

\newcommand{\g}{{{\bmath{g}}}}

\newcommand{\vdot}{{\bmath{\cdot}}}
\newcommand{\grad}{\bmath{\nabla}}

\newcommand{\thth}{\hspace{1.5pt}}

\newcommand\Div{\grad\vdot\thth}

\newcommand{\kpar}{k_{\scriptscriptstyle\parallel}}

\newcommand{\bv}{Brunt-V\"ais\"al\"a}

\renewcommand{\Re}{\mathop{\rm Re}\nolimits}
\renewcommand{\Im}{\mathop{\rm Im}\nolimits}

\newcommand{\mnras}{MNRAS}
\newcommand{\apj}{ApJ}
\newcommand{\apjl}{ApJL}
\newcommand{\apjs}{ApJS}
\newcommand{\aap}{A\&A}

\newcommand{\anap}{AnAp}
\newcommand{\grl}{Geophys. Res. Lett.}
\newcommand{\jgr}{J. Geophys. Res.}

\begin{document}

\title[Radiatively damped gravity waves]{Mode conversion of radiatively damped magnetogravity waves in the solar chromosphere}

\author[M. E. Newington and P. S. Cally]{
Marie E. Newington$^1$\thanks{E-mail: Marie.Newington@monash.edu} and
Paul S. Cally$^{1,2}$\thanks{E-mail: Paul.Cally@monash.edu}\\
$^1$Monash Centre for Astrophysics,
School of Mathematical Sciences,
Monash University, Victoria, Australia 3800\\
$^2$High Altitude Observatory, National Center for Atmospheric Research, P.O. Box 3000, Boulder, CO 80307, USA}

\maketitle

\begin{abstract}
Modelling of adiabatic gravity wave propagation in the solar atmosphere showed that mode conversion to field guided acoustic waves or Alfv\'en waves was possible in the presence of highly inclined magnetic fields. This work aims to extend the previous adiabatic study, exploring the consequences of radiative damping on the propagation and mode conversion of gravity waves in the solar atmosphere.   We model gravity waves in a VAL-C atmosphere, subject to a uniform, and arbitrarily orientated magnetic field, using the Newton cooling approximation for radiatively damped propagation.  The results indicate that the mode conversion pathways identified in the adiabatic study are maintained in the presence of damping.  The wave energy fluxes are highly sensitive to the form of the height dependence of the radiative damping time.  While simulations starting from 0.2 Mm result in modest flux attenuation compared to the adiabatic results, short damping times expected in the low photosphere effectively suppress gravity waves in simulations starting at the base of the photosphere.   It is difficult to reconcile our results and observations of propagating gravity waves with significant energy flux at photospheric heights unless they are generated \emph{in situ}, and even then, why they are observed to be propagating as low as 70 km where gravity waves should be radiatively overdamped.
\end{abstract}

\section{INTRODUCTION}                \label{intro}

Recent multi-height observations of low frequency oscillations in the low solar chromosphere suggest that the energy flux carried by upward propagating gravity waves, with frequencies between 0.7 and 2.1 mHz, comfortably exceeds the co-spatial acoustic wave flux \citep{straus08}.  In our previous paper  \citep{newington10}, (referred to henceforth as paper I), we explored the propagation, reflection and mode conversion of gravity waves in a VAL C  related atmosphere, permeated by uniform, inclined magnetic field.  The significant finding of that study was that in regions of highly inclined magnetic field, gravity waves experience mode conversion to up-going (field-guided) acoustic or Alfv\'en waves.  While acoustic waves are likely to shock before reaching the upper chromosphere, Alfv\'en waves can propagate to greater atmospheric heights, perhaps contributing to the observed coronal Alfv\'enic oscillations \citep{pontieu07,tomczyk07}.  

A limitation of the work presented in paper I was the assumption of adiabatic wave propagation, which is known to be invalid in the photosphere and low chromosphere.   A more realistic investigation of the propagation and mode conversion of gravity waves in the solar atmosphere would include the effects of radiative damping.

Radiatively damped atmospheric gravity waves have been considered by a small number of authors previously, (see \citealt{souffrin66}, \citealt{stix70}, and the exhaustive study by \citealt{mihalas82}), but all of those studies have been purely hydrodynamic in nature, with no applied magnetic field, hence there is no possibility of mode conversion.  Our principal objective in this work is to explore the consequences of radiative damping on the mode conversion of gravity waves, and this requires a magnetohydrodynamic (MHD) treatment.  

In this paper we extend our earlier MHD study, relaxing the assumption of adiabatic wave propagation by incorporating radiative damping using the Newton cooling approximation.  We recognise that this approximation is strictly valid only for optically thin perturbations of a homogeneous, infinite,  and isothermal gas, and that none of these conditions are realised in the photosphere and chromosphere.  The advantage of using this simple treatment is that the same mathematical tools used in paper I can be employed with minor modifications.  The insight gained from this simple model may direct our attention to cases that are interesting enough to warrant a more thorough treatment of the effects of radiation.

The questions central to our investigation are: \textit{How are the mode conversion pathways affected by radiative damping? Do we still get appreciable mode conversion to Alfv\'en waves when damping is present?}    

This paper is organised as follows:  Section 2 presents the atmospheric model and the mathematical tools used in this investigation, namely the damped dispersion relation and numerical solution of the damped linear MHD wave equations.  Section 3  presents the results.  Dispersion diagrams reveal insights into the mode conversion pathways in $z$-$k_z$ phase space, and numerical integration of the wave equations quantify the acoustic and magnetic fluxes.  The conclusions are stated in Section 4.   

\section{MODEL AND EQUATIONS}   \label{model}
This section describes the atmospheric model, the coordinate system and the equations used to generate the results of this paper.  As this work is an extension of the adiabatic investigation of paper I, the reader will be referred to that paper to avoid repetition of material, where appropriate. 

\subsection{Atmospheric model and coordinate system}
We use the same atmospheric  model as employed in paper I -- the \cite{schmitz03} adaptation of the horizontally invariant VAL C model \citep{VAL} up to a height 1.6 Mm above the base of the photosphere.  An isothermal top is appended above 1.6 Mm.  No transition region is included. A uniform inclined magnetic field is imposed upon this atmosphere.  

\begin{figure}
\begin{center}
\includegraphics[width=0.8\hsize]{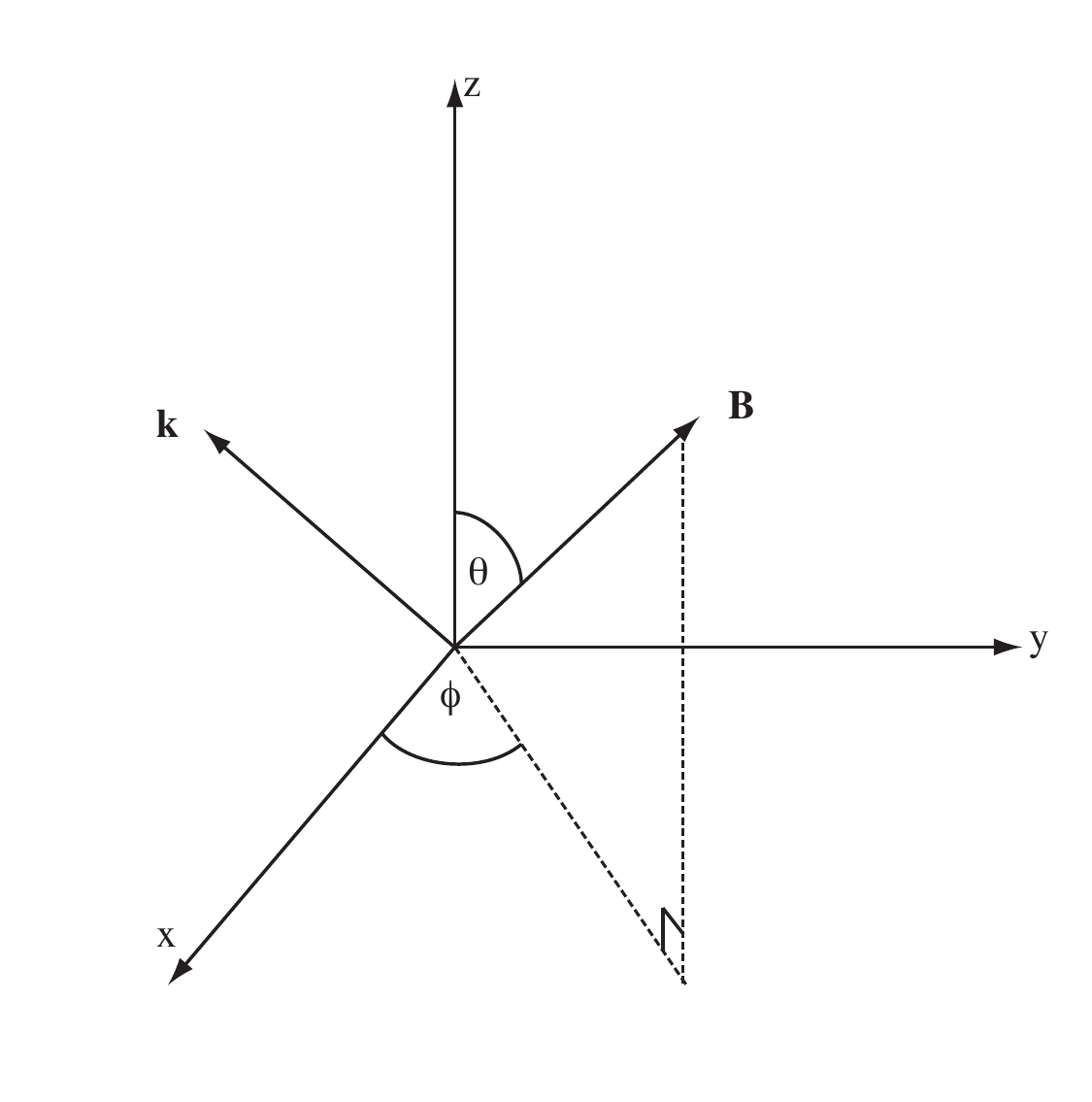}
\caption{The coordinate system and geometry adopted for this analysis.  $\boldsymbol{B}$ is the magnetic field vector and $\boldsymbol{k}$ is the wavevector.  }
\label{fig:geom}
\end{center}
\end{figure}

The origin of the coordinate system is located at the base of the photosphere.  The coordinate system is orientated such that the wavevector ($\boldsymbol{k}$) lies in the  $x$-$z$ plane, and the orientation of the magnetic field ($\boldsymbol{B}$) is described in terms of the inclination from the vertical ($\theta$) and the azimuthal angle ($\phi$)  (see Fig.~\ref{fig:geom}).  We distinguish between two dimensional (2D) and three dimensional (3D) configurations by the angle between the magnetic field and the vertical plane of wave propagation ($\phi$):  2D: $\phi=0$; 3D: $\phi\neq 0$. 

Following paper I, our tools for investigation of radiatively damped gravity wave propagation are the dispersion relation and numerical integration of the linearised MHD wave equations. Ray theory is not directly employed as it is difficult to institute and interpret in the presence of dissipation.  The derivation of the appropriate forms of these equations including Newton cooling are described below.  

\subsection{Mathematical tools incorporating Newton cooling}
\subsubsection{Energy equation and expression for the pressure perturbation}
In the Newton cooling approximation, the radiative damping is parametrized by the radiative relaxation time ($\tau$), assumed a function of height $z$ only.   As $\tau\to\infty$, the motions become adiabatic.  The continuity equation and the momentum equation are unchanged in the Newton cooling approximation, but the energy equation is modified by the inclusion of a term proportional to the temperature perturbation, as follows  (see, for example, \citealt{cally84}):
\begin{equation}
\frac{D p}{Dt}=c^{2}\frac{D\rho}{Dt}-\frac{p_{0}}{\tau}\frac{T_{1}}{T_{0}}\, .
\label{energy}
\end{equation}
Here $\rho$ is the density,  $p$ is the pressure, $T$ is the temperature and $c$ is the sound speed.  The subscripts 0 and 1 denote the background values and Eulerian perturbations respectively.  \\

Upon linearising equation (\ref{energy}), applying the WKBJ approximation, and making use of the perfect gas law, the definition of the adiabatic sound speed $c^2=\gamma p_0/\rho_0$ (where $\gamma$  is the ratio of specific heats), and the magentohydyrostatic balance, the following equation for the pressure perturbation is obtained.  
\begin{equation}
p_{1}=\mu_{1}\rho_{0}g\zeta-\mu_{2}\rho_{0}c^{2}\Div\bxi\,,
\label{presspert}
\end{equation}
where
\begin{align}
\mu_{1}&=(1+\frac{\ri}{\gamma\omega\tau}\frac{c^{2}}{gH})(1+\frac{\ri}{\omega\tau})^{-1}\\
\mu_{2}&=(1+\frac{\ri}{\gamma\omega\tau})(1+\frac{\ri}{\omega\tau})^{-1}
\end{align}
and $g$ is the acceleration due to gravity. $H$ the density scale height and $\bxi=\left(\xi,\eta,\zeta\right)$ is the displacement vector. In the adiabatic limit $\mu_1$ and $\mu_2$ reduce to 1.

\subsubsection{Height dependence of the radiative relaxation time $\tau$}
Although expressions for the radiative damping time have been provided by \cite{spiegel57} and \cite{stix70},  (for  continuum and line emission, respectively), the actual values for the radiative relaxation time in the photosphere and chromosphere are only crudely known.

In this paper it will suffice to adopt the simple linear radiative damping time introduced by \cite{mihalas82} to generate the results in that paper (curve 2): 
\begin{equation}
\tau\left(z\right)=50+(2200/3)z \, ,    \label{tau}
\end{equation}
where $\tau$ is expressed in seconds and $z$ in megameters.

\begin{figure}
\begin{center}
\includegraphics[width=1.0\hsize]{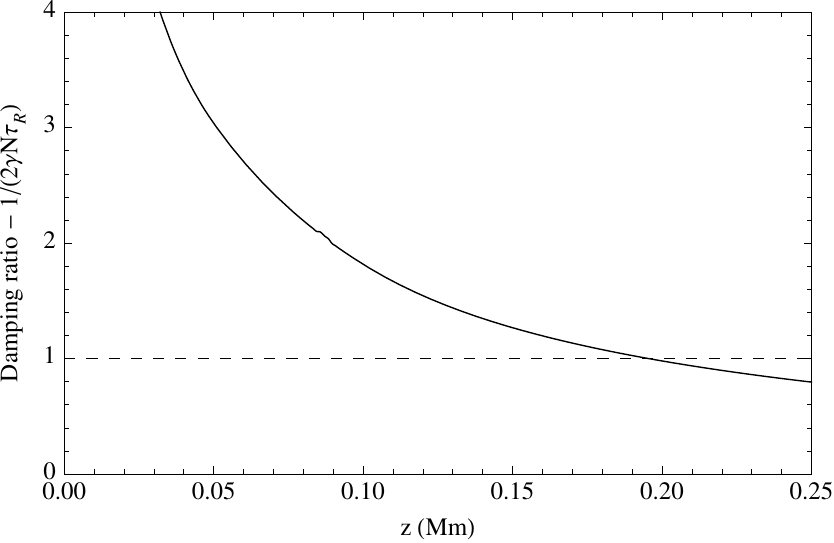}
\caption{The behaviour of the damping ratio as a function of height for the VAL-C atmosphere and the Mihalas \& Toomre (1982) prescription for the radiative damping times.} 
\label{fig:souffrincnd}
\end{center}
\end{figure}

Very short damping times are known to suppress the propagation of gravity waves.  When Newton cooling is included in the energy equation, the velocity of a vertically displaced fluid element may be described in terms of a damped harmonic oscillator, (see, \citealt{souffrin66} and \citealt{bray74}).  This allows identification of the damping ratio as $1/(2\gamma N\tau)$, where $N$ is the {\bv} or buoyancy frequency.   If the damping ratio is greater than one, the motion is overdamped and the fluid parcel will return to the equilibrium position without oscillating.   Applying this condition to their atmosphere, \cite{mihalas82} noted that the wave was overdamped below 0.1 Mm and so they started their simulations from this height.  The atmosphere in this paper has a different height dependence of the {\bv} frequency,  and the location where the damping ratio is 1 is higher at about 0.2 Mm (see fig.~\ref{fig:souffrincnd}). Hence, we expect gravity waves to be underdamped and propagating above 0.2 Mm, and overdamped below this height. 

We also ran simulations using constant relaxation times from 200s to 1 ks to gauge the sensitivity of the wave propagation to the form of  $\tau$, but  (\ref{tau}) was used to generate all the figures in this paper.
 
\subsubsection{Dispersion relation}
\begin{figure}
\begin{center}
\includegraphics[width=1.0\hsize]{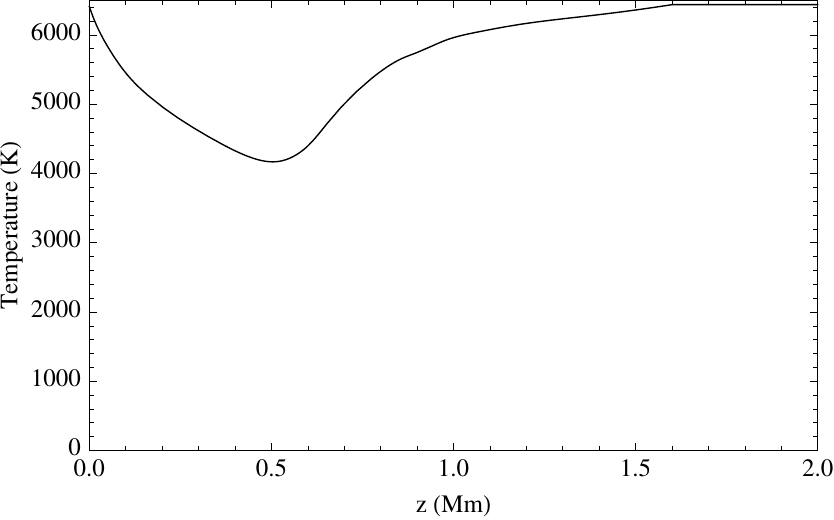}
\caption{Height dependence of temperature in the model atmosphere used here. The VAL-C model is employed  below 1.6 Mm ;  an isothermal layer is appended above 1.6 Mm. }
\label{fig:tempvsz}
\end{center}
\end{figure}

The adiabatic 3D MHD dispersion relation described in \cite{newington10} was derived using a Lagrangian formulation, which is not easily extended to nonadiabatic wave propagation.  However, a dispersion relation for the nonadiabatic case can be readily constructed from the adiabatic relation, if the simplifying assumption of an isothermal atmosphere is adopted.   This is not a bad approximation for the region in question (see figure \ref{fig:tempvsz}).  

In an isothermal atmosphere,  the density scale height is  $H=c^2/\gamma g$,  and so $\mu_1$=1.  The equation for the Eulerian pressure perturbation, equation (\ref{presspert}), then becomes,
\begin{equation}
p_{1}=\rho_{0}g\zeta-\rho_{0}\hat{c}^{2}\Div\bxi \,,  \label{ipresspert}
\end{equation}
where, following \cite{cally84}, we define a new (complex) quantity \footnote{Note that this is equivalent to redefining the ratio of specific heats, as in \cite{bunte94}.}
\begin{equation}
\hat{c}^2=\mu_{2}c^{2} \,.      \label{chat}
\end{equation}

Because the other perturbation equations are unchanged in the Newton cooling approximation, this suggests that for an isothermal atmosphere, the equations describing the damped system will have the identical form to the adiabatic equations,  with the modification that the sound speed squared  $c^2$ is replaced by $\hat{c}^2$.  The damped form of the dispersion relation for an isothermal atmosphere is therefore taken as
\begin{multline}
\omega^2 \hat{\omega}_{\rm c}^2 a_y^2  k_{\rm h}^2 +(\omega^2-a^2\kpar^2)\times{}\\
\left[\omega^4-(a^2+\hat{c}^2)\omega^2 k^2+a^2\hat{c}^2k^2\kpar^2 
\right.\\ \left.{}
+ \hat{c}^2\hat{N}^2 k_{\rm h}^2
-(\omega^2-a_z^2k^2) \hat{\omega}_{\rm c}^2\right]=0,   \label{DF}
\end{multline}
where $a_z$ is the vertical component of the Alfv\'en velocity and $a_y$ is the
component perpendicular to the plane containing $\k$ and $\g$.  $\hat{N}$ is defined by $\hat N^2=g/H-g^2/\hat{c}^2$, $ \hat{\omega}_{\rm c}=\hat{c}/2H$, and $ k_{\rm h}$ is the horizontal component of the wavevector.

Although \emph{ad hoc}, the above dispersion relation will prove useful in Section \ref{disp} in understanding the connectivity of  gravity waves to field guided acoustic (slow) waves or Alfv\'en waves in the upper atmosphere where $a\gg c$.  (See \cite{newington10} for the low-$\beta$ asymptotic solutions of the 3D MHD dispersion relation.)  As in paper I, we corroborate the dispersion diagrams with numerical solutions of the linearized wave equations, the derivation of which \emph{does not} assume an isothermal atmosphere.  This is described in the following section.

\subsubsection{Numerical solution of linearised, damped MHD equations}
The linearised, MHD equations incorporating Newton cooling are obtained from using the (non-isothermal) expression for the pressure perturbation (\ref{presspert}), in the momentum equation, and eliminating  the density by means of the continuity equation.  Expressed in terms of the cartesian components of the Lagrangian displacement vector,  $\bxi=\left(\xi,\eta,\zeta\right)$ the equations are as follows: 
 \begin{multline}
\omega^{2}\xi-k_x\left(\ri g\zeta\mu_{1}+ c^{2}\mu_{2}\left(k_x\xi-\ri\zeta^{\prime}\right)\right)\\
+ a^{2}\left[k_x^{2}\cos\theta\cos\phi\sin\theta\,\zeta+k_x^{2}\cos\phi\sin^{2}\theta\sin\phi\,\eta\right. \\ 
\left.- k_x^{2}\cos^{2}\theta\,\xi - k_x^{2}\sin^{2}\theta\sin^{2}\phi\,\xi+ik_x\sin^{2}\theta\sin^{2}\phi\,\zeta'\right.\\
\left.-\ri\,k_x\cos\theta\sin\theta\sin\phi\,\eta'- \cos\theta\cos\phi\sin\theta\,\zeta^{\prime\prime}+\cos^{2}\theta\,\xi''\right]=0\,,
\end{multline}
\begin{multline}
-\frac{c^{2}}{H}\mu_{2}\left(\ri k_x \,\xi+\zeta^{\prime}\right)\\
+g\left(\left(1-\mu_{1}\right)\zeta^{\prime}+\ri k_x\,\xi\right)-\frac{g}{H}\zeta\left(1-\mu_{1}+H\mu_{1}^{\prime}\right)\\
+ \left(\omega^{2}-k_x^{2}a^{2}\cos^{2}\phi\,\sin^{2}\theta -\mu_1 g'\right)\zeta+\ri k_x\,\xi\,\mu_{2}c^{2\prime}+\mu_{2}c^{2\prime}\zeta^{\prime}\\+\ri k_x\,c^{2}\mu_{2}\xi^{\prime}+\ri k_x\,c^{2}\xi\mu_{2}^{\prime}+c^{2}\zeta^{\prime}\mu_{2}^{\prime}+c^{2}\mu_{2}\zeta^{\prime\prime}\\
+a^{2}\sin\theta\left[k_x^{2}\cos\phi\cos\theta\,\xi -\cos\theta\left(\sin\phi\,\eta^{\prime\prime}+\cos\phi\,\xi^{\prime\prime}\right)\right.\\
+\left.\sin\theta\left(\ri k_x\sin\phi\left(\sin\phi\,\xi'-\cos\phi\,\eta'\right)+\zeta^{\prime\prime}\right)\right]=0\,,
\end{multline}
\begin{multline}
\left(\omega^{2}-k_x^{2}a^{2}\,\cos^{2}\phi
\,\sin^{2}\theta\right)\eta\\
+a^{2}\left[\sin\theta\left(k_x\cos\phi\sin\theta\sin\phi\left(k_x\xi-\ri\zeta^{\prime}\right)\right.\right.\\
+\left.\left.\cos\theta\left(2\ri\, k_x\cos\phi\,\eta^{\prime}-\sin\phi\left(\ri k_x\,\xi^{\prime}+\zeta^{\prime\prime}\right)\right)\right)+\cos^{2}\theta\,\eta^{\prime\prime}\right]=0
\end{multline}
The Alfv\'en speed is denoted by $a$.   

The purely horizontal case $\theta=90^\circ$ is not examined here as it is singular in nature, with the governing equations reduced in order resulting in `critical levels' producing resonant absorption at the Alfv\'en and cusp resonances \citep{cally84}. This `absorption' may in fact be interpreted as a mode conversion, where the absorption coefficient is continuous with $\theta$ as it reaches $90^\circ$ (as found by \cite{cally11} in the case of the Alfv\'en resonance), so physically there is nothing particularly special about horizontal field despite its mathematical peculiarity. The singularities in the solutions are regularised in practice due to solar atmospheric waves not being strictly monochromatic.

As in paper I, a driven wave scenario is envisaged, where monochromatic upward propagating (in terms of group velocity) gravity or acoustic waves are assumed to be excited at the bottom of our region of interest. In the interest of isolating the effects of mode conversion of acoustic and gravity waves during their propagation through the atmosphere, the condition that no slow magneotacoustic waves or Alfv\'en  waves entered the photosphere from below is imposed. Conversely, no waves of any sort are allowed to enter from the top.  

The choice of a boundary value problem means that we considered the spatial damping of these waves (see \cite{souffrin72}). Temporal invariance of the atmospheric model considered means that the wave frequency $\omega$ is constant; this is determined by the driving force and is real. Horizontal invariance ensures constancy of the horizontal component of the wave number $k_x$ during the wave's propagation.  Damping results in the $z$ component of the wavenumber $k_z$ being complex.  The magnitude of the imaginary component of $k_z$ increases with the strength of the damping.  

Solution of the equations requires an arbitrary normalisation condition be applied.  In paper I this was chosen to be normalisation of the vertical velocity, $w$ to 1 km$^{-1}$ at the base of the photosphere ($z=0$).  For most of the results presented in section \ref{results}, we chose to normalise $w$ to 1 km$^{-1}$ at the base of  the underdamped region (0.2 Mm), but we also ran simulations where the wave was normalised at the base of the photosphere (0 Mm) to observe the effect of overdamping on the transmitted fluxes.  

The motion is assumed adiabatic ($\tau\to\infty$) at heights above 1.6 Mm, where the radiative decay time becomes very long, and below the height of application of the normalisation condition.  This simplification allows us to apply the same boundary conditions as in paper I, without modification.  The reader is referred to that paper for details. 

The acoustic and magnetic fluxes were calculated in the same manner as in paper I.

\section{RESULTS AND DISCUSSION}   \label{results}

\subsection{Dispersion diagrams}  \label{disp}
\begin{figure*}
\begin{center}
\includegraphics[width=\hsize]{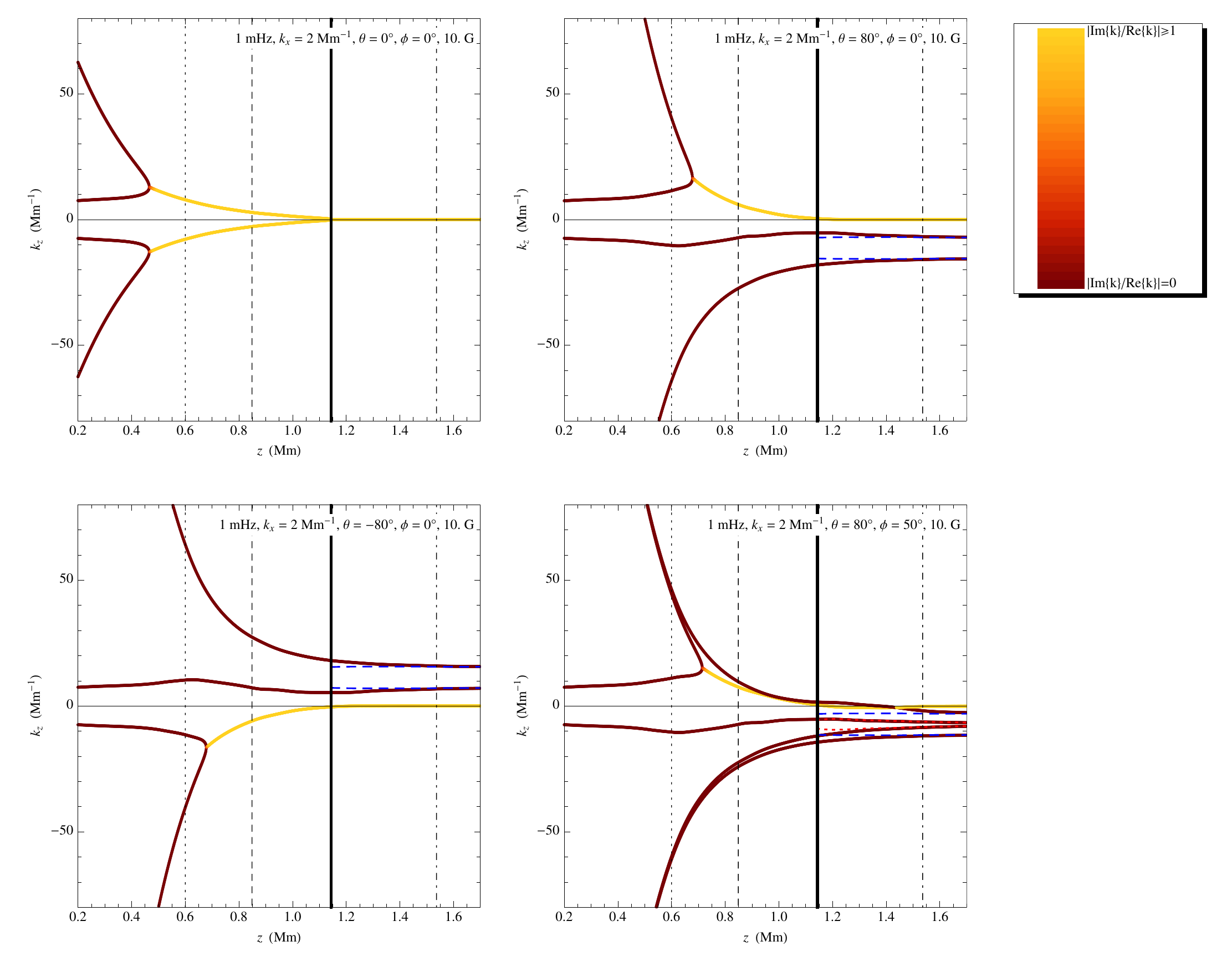}
\caption{Dispersion diagrams for adiabatic propagation of  waves of frequency 1 mHz and horizontal wavenumber  $k_x$=2 Mm$^{-1}$, subject to a 10G magnetic field, included for the purpose of comparison to the radiatively damped results.   In this dispersion diagram and those that follow, the vertical lines indicate the value of the ratio $a^2/c^2$ at various heights: solid line --  $a^2/c^2=1$; dashed line --  $a^2/c^2=0.1$; dotted line --  $a^2/c^2=0.01$; dot-dashed line --  $a^2/c^2=10$. The curve is coloured according to the damping per wavelength -- the larger the magnitude, the greater the damping, and the more yellow the curve.  The asymptotic solutions (detailed in paper I) corresponding to  the field-guided acoustic waves and Alfv\'en waves are shown as blue dashed, and red dotted curves, respectively.   The bottom branch corresponds to the up-going gravity wave.  The four panels differ in the field orientations --  Top left panel: vertical field.  The gravity wave is reflected as a slow wave; top right panel: 2D inclined field where the field lines are closely aligned with the direction of wave propagation.  The gravity wave penetrates the equipartition level and undergoes mode conversion to a field-aligned acoustic wave;  Bottom left panel: 2D inclined, anti-aligned  field.  When the attack angle (the angle between the direction of wave propagation and the magnetic field) is too large, the gravity wave reflects as a slow wave;  Bottom right panel: 3D inclined  field.   Here the gravity wave connects to the Alfv\'en wave solution.  This figure presents the same results as figure 4 of paper I, but with the z axis starting at 0.2 Mm.  Note also that the plots shown here include  a predominantly evanescent mode (yellow) that is absent in the original figure. } 
\label{fig:dispgridad}
\end{center}
\end{figure*}

\begin{figure*}
\begin{center}
\includegraphics[width=\hsize]{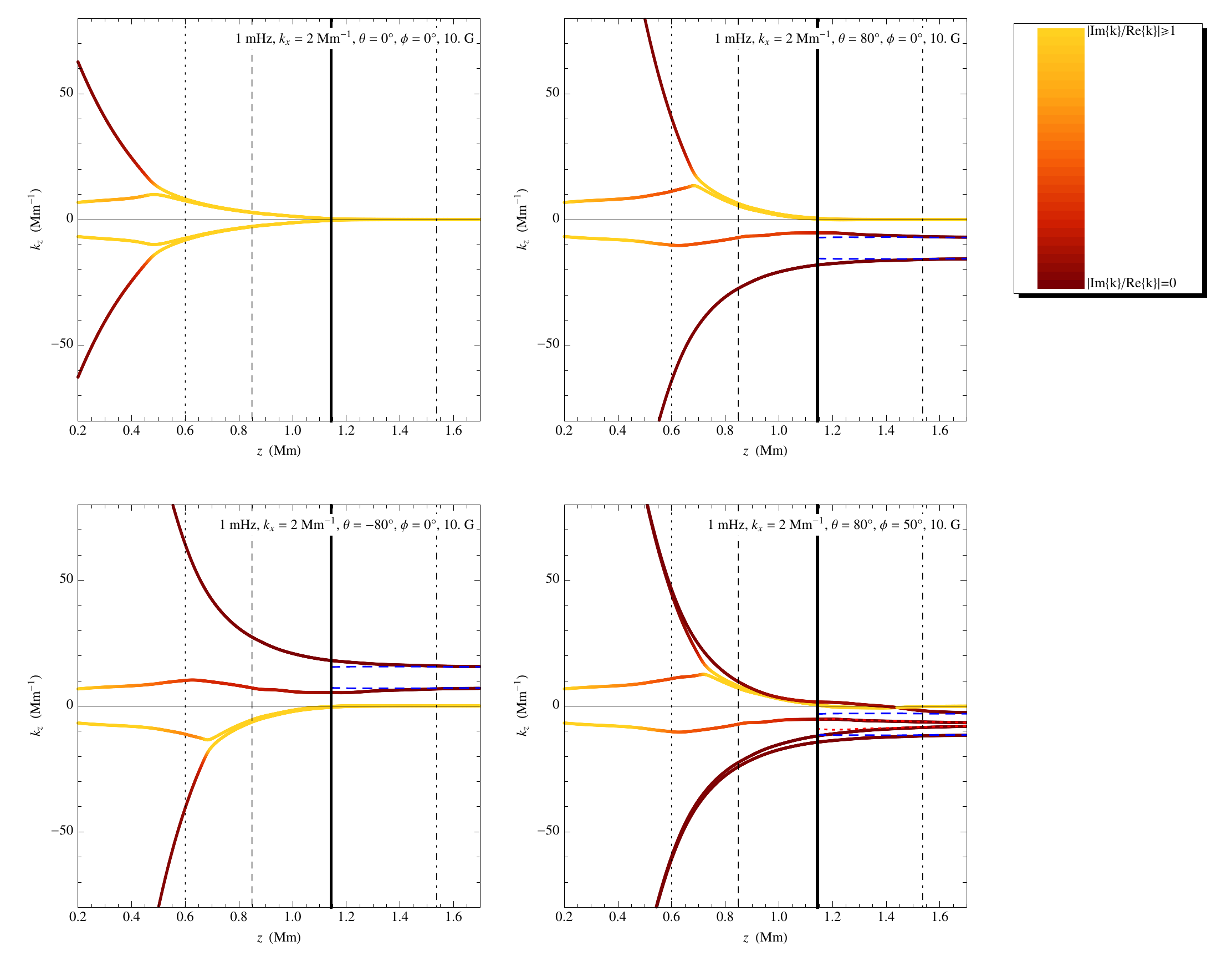}
\caption{The same gravity wave propagation scenarios as in figure \ref{fig:dispgridad}, except that in this case the propagation is subject to radiative damping.  The colour of the curves indicates the damping per wavelength.  Damping is greatest for the gravity wave in the photosphere, with the energy losses decreasing as the height increases.  (Recall that the yellow curves with $k_z$ tending to zero high up are the predominantly evanescent modes present in the adiabatic case.)   Comparison with the adiabatic results (figure \ref{fig:dispgridad}) illustrates that the mode conversion pathways found from the adiabatic simulations, are preserved in the presence of damping.  }
\label{fig:dispgrid_mt}
\end{center}
\end{figure*}

We present diagnostic dispersion diagrams for four different orientations of the magnetic field.  Figure \ref{fig:dispgridad} presents the adiabatic results.  This is the same diagram as figure 4 in paper I, except that in this paper the z axis has been truncated at the lower boundary of 0.2 Mm, and an essentially evanescent branch is included (yellow).  Figure \ref{fig:dispgrid_mt} displays results obtained for damped wave propagation, using the isothermal dispersion relation,  (\ref{DF}).   Recall that the lower branch of the dispersion curve corresponds to the up-going gravity wave. The asymptotic solutions for the field guided acoustic wave and the Alfv\'en wave are indicated  in the region above the equipartition level, by the dashed blue and dotted red curves, respectively. 

In both figures, the curves are colour coded according to the damping per wavelength ($\Im{\{k\}}/\Re{\{k\}}$), where the more yellow the curve, the larger the value, and the heavier the damping.  In the adiabatic results, (fig. \ref{fig:dispgridad}), the yellow curves represent a predominantly evanescent mode that is absent from the figure in paper I.  We draw attention to it so that its counterpart can be identified in the damped figure, but we are not concerned with this mode in this paper.  The colour coding in figure \ref{fig:dispgrid_mt} shows that the up-going gravity wave is most heavily damped lower in the atmosphere, which is to be expected from the linear form of $\tau$.   

The dispersion diagrams imply that the behaviours found in the adiabatic case, concerning the wave propagation behaviour with the field orientation and the character of the dominant mode conversion, are largely unaffected by the damping.  Comparison of figures  \ref{fig:dispgridad} and \ref{fig:dispgrid_mt} show that the wave paths in $z-k_z$ space are preserved when damping is included.  This implies that the conclusions drawn in paper I about gravity wave behaviour with various field orientations are also valid when the gravity wave experiences radiative damping.  The connectivities to the asymptotic solutions (which are indicative of the dominant mode conversion) are preserved despite the introduction of damping.  Numerical integration is used to confirm the dominant mode conversions in the following subsection. 

The paths in  $z$--$k_z$ space are largely unchanged when constant values of $\tau$ (ranging from 200s to 1ks) are used instead of the linear form (\ref{tau}). The strength of the damping along the paths varies with each form of $\tau$.  Of course, the net flux emerging at the top depends on this damping along the whole path, and is addressed in the next subsection. The results presented there also justify the crude assumptions made in our dispersion analysis \emph{ex post facto}.


\subsection{Transmitted fluxes}
\begin{figure}
\begin{center}
\includegraphics[width=1.0\hsize]{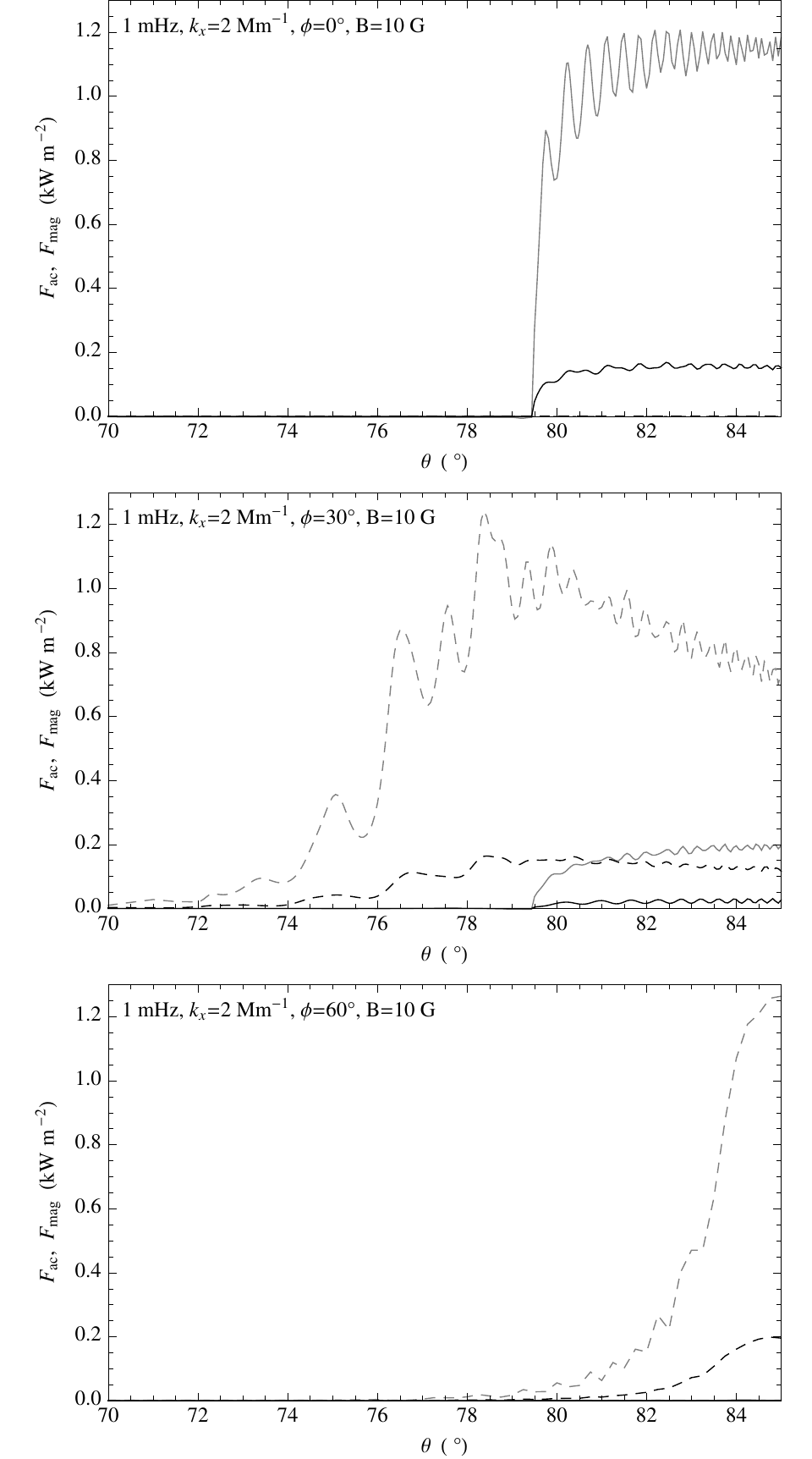}
\caption{Acoustic (full) and magnetic (dashed) wave-energy flux ($\rm kW\,m^{-2}$) as functions of magnetic field inclination $\theta$ and three different orientations $\phi$ (0$^\circ$, 30$^\circ$ and 60$^\circ$, respectively), for B=10 G and 1 mHz waves with $k_x=2$ Mm$^{-1}$.   The black curves indicate the results obtained for wave propagation subjected to damping, while the grey curves are the results of adiabatic simulations for comparison. We apply the normalisation condition at 0.2 Mm.  There the vertical velocity $w$ is normalised to 1 km s$^{-1}$.  }
\label{fig:fluxes_mt2}
\end{center}
\end{figure}

\begin{figure*}
\begin{center}
\includegraphics[width=1.0\hsize]{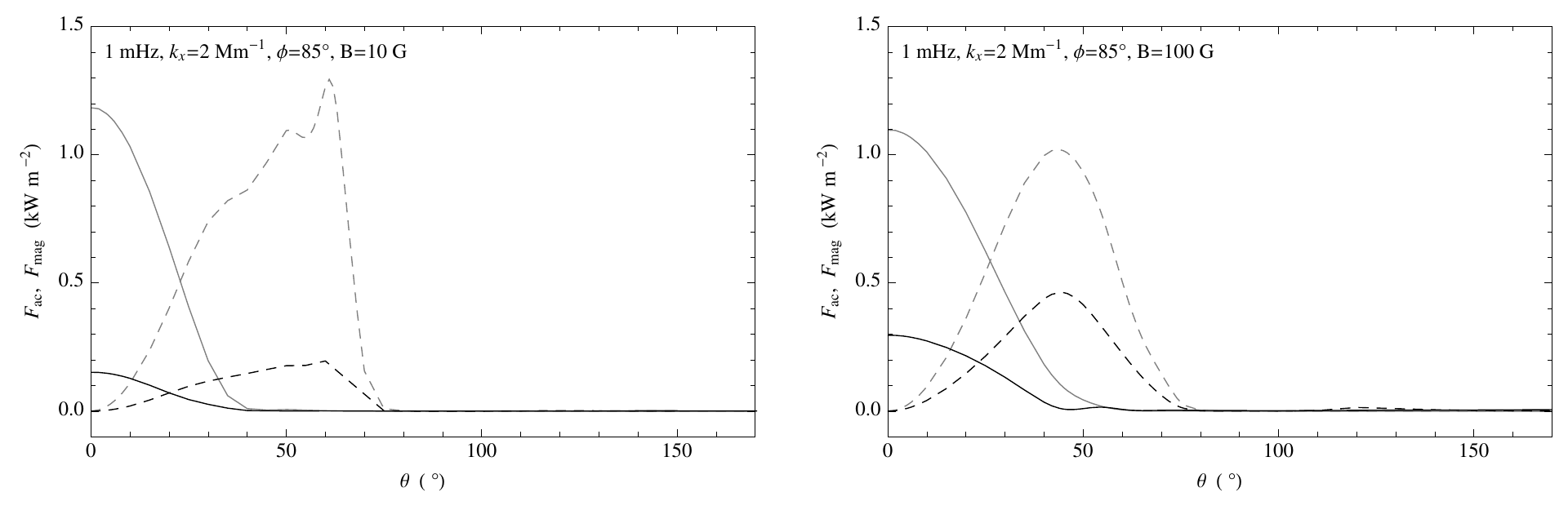}
\caption{Acoustic (full) and magnetic (dashed) wave-energy fluxes (kW m$^{-2}$) as functions of magnetic field orientation $\phi$ with inclination $85^\circ$.  Results of damped and adiabatic simulations are the black and grey curves, respectively. }
\label{fig:fluxes_phi_mt2}
\end{center}
\end{figure*}

\begin{figure}
\begin{center}
\includegraphics[width=1.0\hsize]{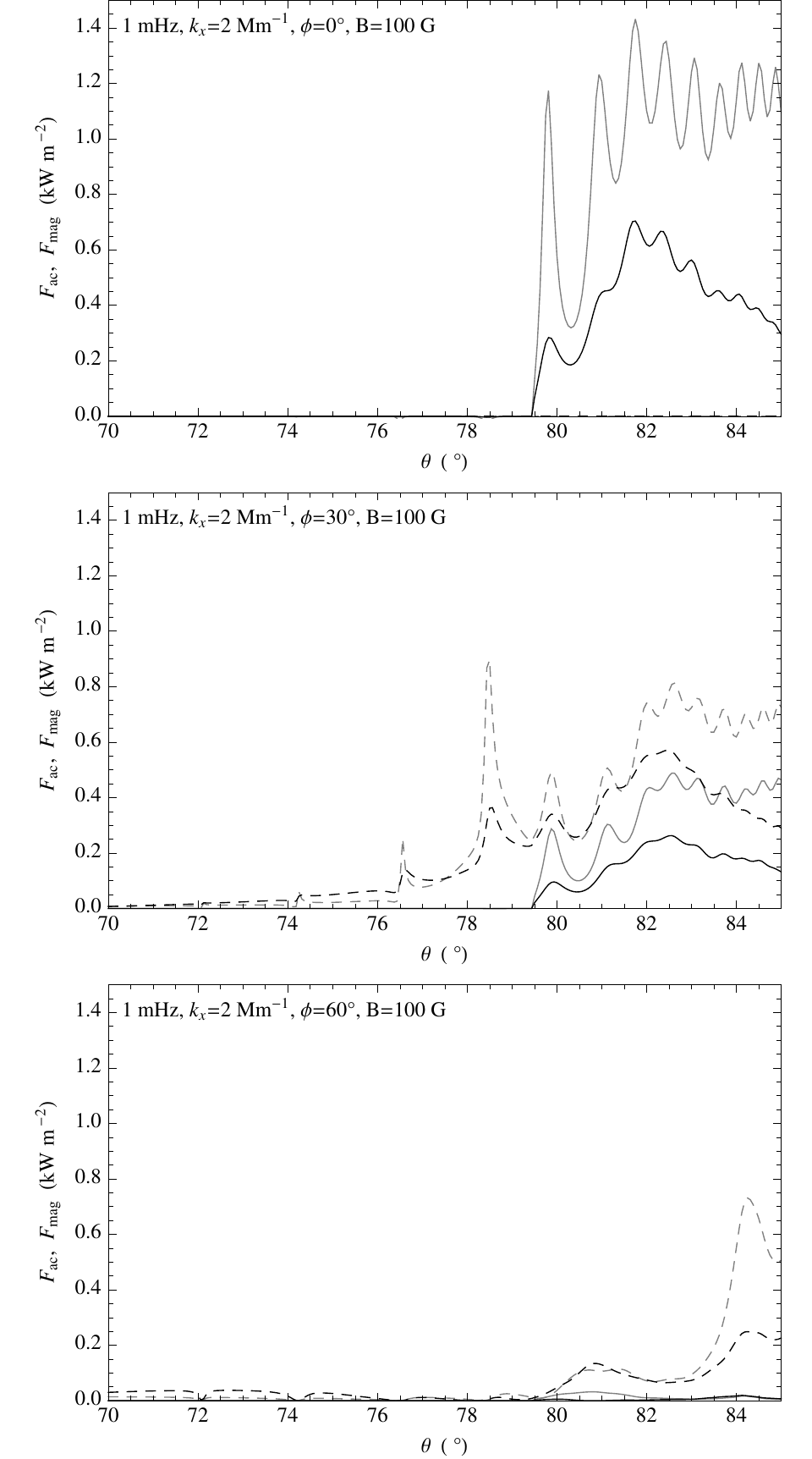}
\caption{As for Figure \ref{fig:fluxes_mt2}, but with B=100G }
\label{fig:fluxes_mt2_100G}
\end{center}
\end{figure}

In all cases the radiative damping results in a reduction in magnitude of the measured acoustic and magnetic fluxes with respect to the adiabatic results.  Performing the integrations with both linear and constant forms of $\tau$ revealed that the magnitude of the flux reduction in the damped simulations is very sensitive to the form of the radiative damping used.  Our investigations allow us to make qualitative statements  regarding the effects of magnetic field strength and orientation, the effect of frequency, and the impact of the depth of the application of the normalisation condition, and we illustrate them with the figures obtained using (\ref{tau}). However, bear in mind that the actual values of the measured fluxes in the figures are contingent on the use of the \cite{mihalas82} form of the radiative damping time.  

\subsubsection{Effect of magnetic field orientation}
 Figure \ref{fig:fluxes_mt2}, and the left hand panel of Figure \ref{fig:fluxes_phi_mt2} show the behaviours of the magnetic and acoustic fluxes as functions of the inclination $\theta$ and azimuthal angle $\phi$ for a gravity wave with frequency 1 mHz and horizontal wavenumber 2 Mm$^{-1}$, in an atmosphere with an imposed magnetic field of 10G.   Figure  \ref{fig:fluxes_mt2} shows the behaviour at three different azmiuthal angles.  As expected from the dispersion diagrams, apart from the obvious reduction in flux magnitude, the behaviour of the fluxes with magnetic field orientation in the damped simulations is similar to that observed in the adiabatic case.   Notably, the angles at which the ramp effect turns on and the mode conversion takes place are the same in the adiabatic and damped cases.   This was observed in simulations with constant $\tau$ too.

\subsubsection{Effect of magnetic field strength}

Figure \ref{fig:fluxes_mt2_100G} and the righthand panel of Figure \ref{fig:fluxes_phi_mt2} show the results calculated with a magnetic field strength of 100G.  Comparison with the results generated with a weaker field strength of 10G  (fig. \ref{fig:fluxes_mt2}), reveals that the waves experience less flux attenuation in the high field strength cases. Although magnetic pressure and tension play a larger role in driving waves at higher field strengths, this result is not obvious \emph{a priori}.


\subsubsection{Effect of frequency }
\begin{figure}
\begin{center}
\includegraphics[width=1.0\hsize]{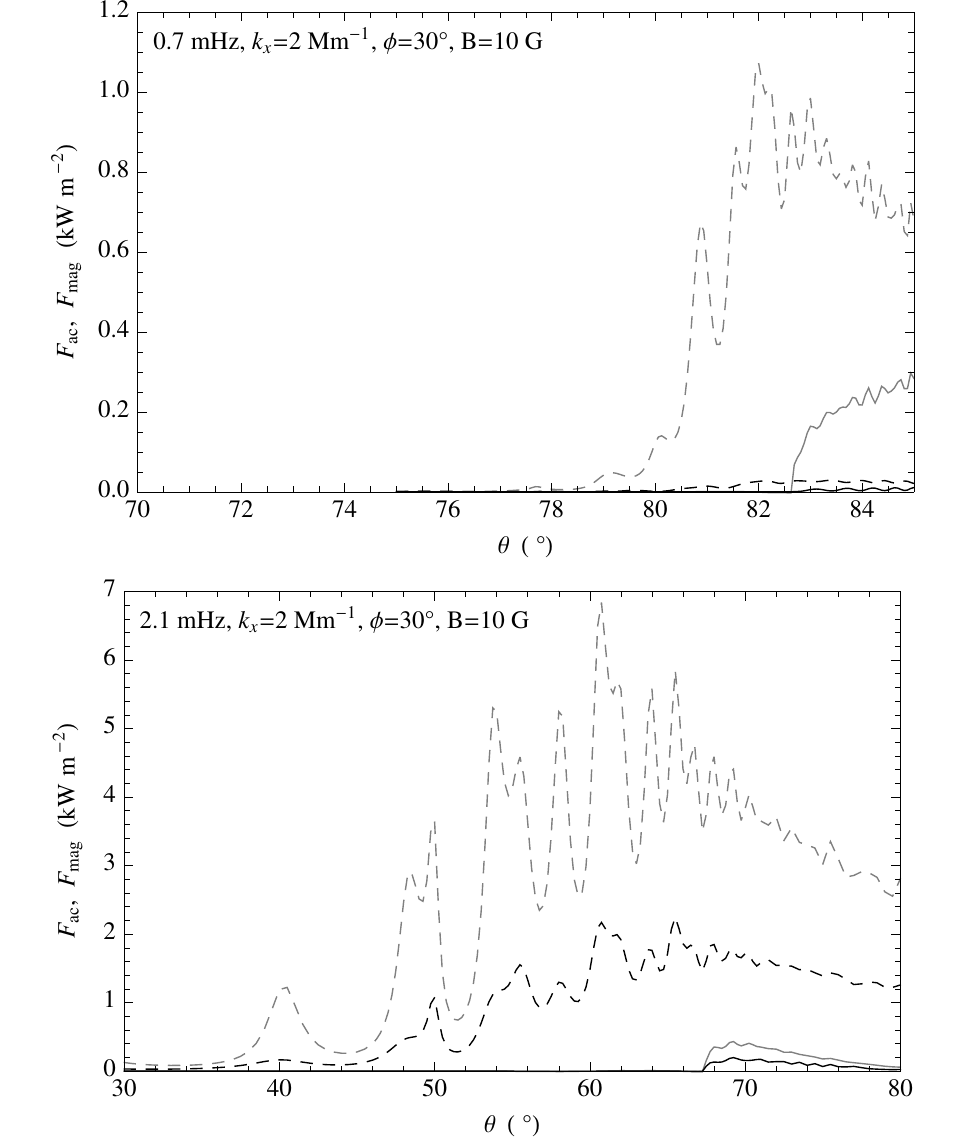}
\caption{Acoustic (full) and magnetic (dashed) damped (black) and adiabatic (grey) wave-energy fluxes (kW m$^{-2}$) as functions of magnetic field inclination $\theta$ with $\phi=30^\circ$ for  B=10 G, $k_x=2$ Mm$^{-1}$ for waves of two different frequencies.  Top panel: 0.7 mHz;  Lower panel: 2.1 mHz.  Note the different $\theta$ scales on the two graphs.  }
\label{fig:fluxes_freq}
\end{center}
\end{figure}

Figure \ref{fig:fluxes_freq} shows the results calculated for two different frequencies.  The top panel is for waves with frequency 0.7 mhz, and the lower panel is for 2.1 mHz.  This shows that the low frequency waves experience greater flux attenuation than waves of higher frequency. This is to be expected since shorter period waves naturally lose less energy radiatively at fixed $\tau$.

\subsubsection{Effect of the depth of application of normalisation condition}

\begin{figure}
\begin{center}
\includegraphics[width=1.0\hsize]{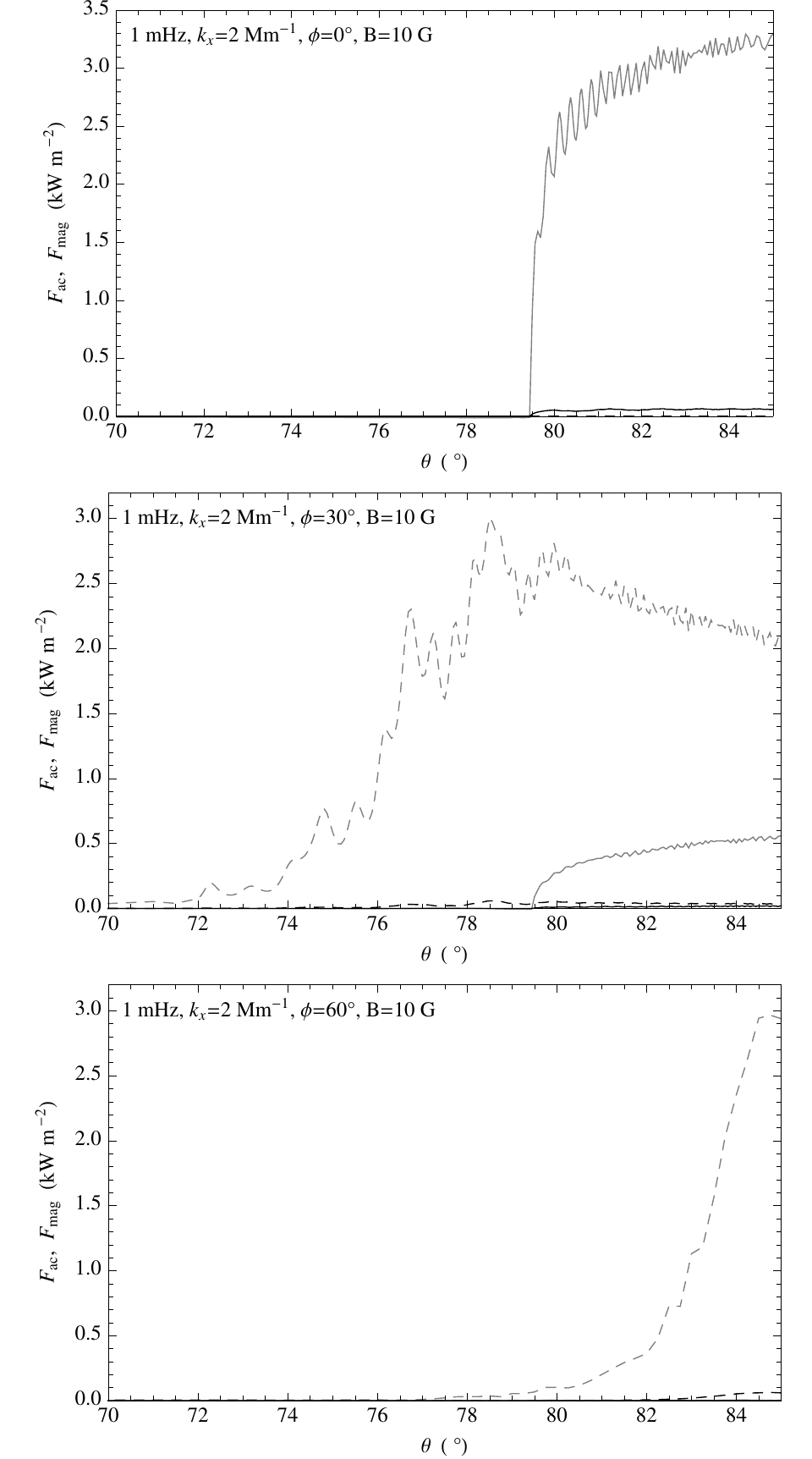}
\caption{As for \ref{fig:fluxes_mt2}, but with the normalisation condition now applied deeper at z=0.  Very little flux is measured at the top when the gravity wave is damped.  With the chosen form of $\tau(z)$, the short damping times in the low photosphere very effectively  mitigate gravity wave progression. }
\label{fig:subphotgen}
\end{center}
\end{figure}

Figure \ref{fig:subphotgen}  shows the fluxes as a function of magnetic field inclination generated with the normalisation condition applied at the base of the photosphere at $z=0$ Mm.  In these simulations, the gravity wave is effectively extinguished, as was predicted by \cite{mihalas82}.

\section{CONCLUSIONS}     \label{conclusions}
The main conclusions we draw from the results of our investigation are:

\begin{enumerate}
\item The primary effect of radiative damping in our model is a decrease in the measured wave energy fluxes obtained, with respect to the adiabatic results.  The extent of the flux attenuation is highly sensitive to the radiative damping times.  

\item Damping does not have a great effect on the mode conversion pathways in the $z-k_z$ plane. Specifically, direct coupling to Alfv\'enic disturbances higher up in the chromosphere seems to survive (in diminished form) the introduction of radiative losses.

\item Short damping times very effectively extinguish gravity waves.  This  is most likely to be a problem in the low photosphere, where radiative damping times are expected to be far shorter than wave periods, rendering the waves overdamped.  Reconciling observations of propagating photospheric gravity waves \citep{straus08,stodilka08} with this seems an impossibility unless they are being generated there \emph{in situ}.\footnote{In Earth and planetary atmospheres, it is known that primary gravity waves can produce secondary gravity waves by nonlinear processes such as wave breaking and the production of turbulent cascades \citep{lane06}, and non-linear wave-wave interactions \citep{dong88, huang07}. Such processes do not seem to be available to us in the low solar atmosphere.}   \cite{komm91} suggest that gravity waves are generated by decay of granular motions in the mid photosphere. \cite{stodilka08} found that overshooting convection starts at around $100$ km and apparently detected gravity waves above this height.  Gravity wave excitation by penetrative convection (albeit in the context of the convection zone/radiative core interface) has been modelled numerically \citep{din03,din05} and found to be effective.

\item Even accepting \emph{in situ} gravity wave generation, it is not clear that \cite{straus08} see \emph{propagating} gravity waves as low as 70 km, where they should be substantially overdamped, notwithstanding the rather arbitrary form of our radiative decay time $\tau(z)$. Indeed, their own numerical simulations using the (nonmagnetic) radiation hydrodynamics \textsf{CO$^\mathsf{5}$BOLD} code, which has a much more sophisticated treatment of radiation than we have instituted, is dominated by convection at $\sim70$ km, with a negligible gravity wave flux there (Straus, private communication). Gravity waves are progressively generated as height increases, reaching their full strength at around 300 km, with the gravity wave being `free' only above this. This justifies our decision to model gravity waves in $z>200$ km; the region below 200--300 km is where the waves are driven.

\end{enumerate}

\section*{ACKNOWLEDGMENT}
The authors would like to gratefully acknowledge Stuart Jefferies and Thomas Straus for helpful discussion and clarification of details of their simulations.

The National Center for Atmospheric Research is sponsored by the National Science Foundation.




\end{document}